\documentclass{iopart}
\usepackage{iopams,graphicx,bm,cite,amsbsy}
\newcommand{\be}{\begin{equation}}
\newcommand{\ee}{\end{equation}}
\newcommand{\bg}{\bm{\gamma}}
\renewcommand{\br}{\bm{\sigma}}
\renewcommand\v{{\mathbf v}}
\newcommand\x{{\mathbf x}}
\renewcommand\P{\widetilde{P}}
\renewcommand\e{{\mathbf e}}
\renewcommand\i{{\rm i}}

\newcommand{\vu}{{\mathbf u}}

\begin{document}
\title{Full counting statistics of chaotic cavities from classical
action correlations}
\author{G Berkolaiko$^1$, J M Harrison$^{1,2}$, M Novaes$^3$}
\address{$^1$ Department of Mathematics, Texas A\&M University,
College Station, TX 77843-3368, USA\\$^2$ Department of Mathematics,
Baylor University, Waco, TX 76798-7328, USA\\$^3$ School of Mathematics,
University of Bristol, Bristol BS8 1TW, UK}

\begin{abstract}

We present a trajectory-based semiclassical calculation of the full
counting statistics of quantum transport through chaotic cavities,
in the regime of many open channels. Our method to obtain the $m$th
moment of the density of transmission eigenvalues requires two
correlated sets of $m$ classical trajectories, therefore
generalizing previous works on conductance and shot noise. The
semiclassical results agree, for all values of $m$, with the
corresponding predictions from random matrix theory.

\end{abstract}


\section{Introduction}

Phase-coherent electron transport through ballistic quantum dots display a
number of universal properties if the corresponding classical dynamics is
chaotic \cite{ferry}. These are well described by random matrix theory (RMT),
in which the system's scattering matrix is assumed to be a random element of
the appropriate ensemble \cite{rmp69cwjb1997}, {\it i.e.} to be random unitary
or unitary symmetric depending on whether time reversal symmetry is absent or
present, respectively. RMT is therefore concerned with the average behaviour of
a collection of different systems within a given universality class determined
solely by the existing symmetry. On the other hand, properties of generic
individual chaotic systems are expected to agree with the predictions of this
theory provided a local energy average is considered, around a classically
small but quantum mechanically large energy window. Rigorously deriving this
connection between chaos and universality is one of the main challenges of the
semiclassical trajectory-based approach to mesoscopic transport
\cite{c3hub1993,prl89kr2002,njp9sm2007}.

As in the case of spectral statistics of closed chaotic systems \cite{form},
the main ingredient from the classical dynamics is the existence of
correlations between long trajectories. They organize themselves into families
according to their action, and the elements of a family differ among
themselves only by their behaviour in small regions (much smaller than their
total length) in which some of them have crossings while others have
anticrossings. The correlations induced by the existence of crossings is
responsible for the emergence of universal quantum properties. This approach
has been successful in reproducing RMT results for the average conductance
\cite{prl89kr2002,prl95sh2005}, shot noise
\cite{prl91hs2003,prl96rsw2006,jpa39pb2006}, time delay \cite{kuipers}, and
other properties \cite{njp9sm2007}. All these calculations have a natural
perturbative structure in which $1/N$, the inverse number of open quantum
channels, plays the role of a small parameter.

In this work we advance this line of investigations by obtaining, using the
semiclassical approximation and classical correlations, the full counting
statistics of chaotic cavities: the complete set of moments of the density of
transmission eigenvalues. Physically, this encodes information about the
statistics of the electric current through the system
\cite{noise,jsm2005,prl91br2003}, viewed as a random time signal. The first
two such moments are related to the average conductance and shot noise. Our
result is restricted to leading order in $1/N$, and to this extent we conclude
that all linear statistical information contained in the RMT of quantum
chaotic transport is also contained in the semiclassical approximation.

In order to help put our results in perspective, let us draw an analogy with
closed systems. By far the most popular quantity to be calculated in that case
is the spectral form factor, the Fourier transform of the $2$-point correlation
function. The calculation requires taking into account contributions due to
pairs of periodic orbits. By contrast, in the present work we consider two sets
of trajectories with $m$ elements each for all values of $m$. In a closed
system this would correspond to obtaining all $m$-point correlation functions.

\section{Counting statistics of chaotic transport}

Quantum transport is governed by the transmission matrix $t$, or equivalently
the hermitian matrix $tt^\dag$. This matrix has a set $\{T_1,\ldots,T_n\}$ of
$n={\rm min}\{N_1,N_2\}$ non-zero transmission eigenvalues, where $N_1$ and
$N_2$ are the number of open channels in the incoming and outgoing leads,
respectively. We consider a chaotic cavity with typical dwell time $\tau_D$,
Lyapunov exponent $\lambda_L$ and linear size $L$. Together with the Fermi
wavelength $\lambda_F$, these last two quantities define the Ehrenfest time
$\tau_E=\lambda_L^{-1}\log(L/\lambda_F)$, roughly the time it takes for an
initially localized wave packet to spread to the size of the system. For times
much longer than $\tau_E$ one can expect any initial state to become
effectively equidistributed.

The regime we are interested in is the semiclassical limit $\lambda_F\ll L$,
when there are many open channels, $N_1,N_2\gg 1$. However, we must approach
this limit in such a way that $\tau_D$, which is a classical time scale,
satisfies $\tau_D\gg\tau_E$. Since $\tau_E$ grows only very slowly as
$\lambda_F\to 0$, one may think of a suitable simultaneous shrinking of the width of the
leads. This is the regime in which universality due to chaos is expected (for
studies considering the situation when $\tau_E\gtrsim\tau_D$, see
\cite{Ehrenfest} and references therein), and RMT predicts that the
transmission eigenvalues are distributed in the interval $[1-4\xi,1]$ with
average density given by \cite{rmp69cwjb1997,rho2}
\begin{equation}
  \label{density}
  \rho(T)=\frac{N}{2\pi T}\sqrt{\frac{4\xi}{1-T}-1}.
\end{equation}
Here $N=N_1+N_2$ is the total number of channels and the variable $\xi$ is
defined as
\begin{equation}
  \xi=\frac{N_1N_2}{N^2}.
\end{equation}
This result is valid to leading order in $N^{-1}$, and is the same for all
universality classes. The presence or absence of time-reversal symmetry is only
felt in higher-order terms, sometimes called `weak-localization' corrections.

The function $\rho(T)$ can be characterized by its moments,
\begin{equation}
  \label{M1}
  M_m=\int\rho(T)T^mdT=\langle {\rm Tr}[(tt^\dag)^m]\rangle,
\end{equation}
where the brackets denote an average over the corresponding random matrix
ensemble.  The first two moments are related to the conductance
($\propto M_1$) and to the shot noise ($\propto M_1-M_2$).  For general $m$,
the RMT moments were calculated explicitly \cite{prb75mn2007},
\begin{equation}
  \label{rmt}
  M_m=N\xi\sum_{p=0}^{m-1}{m-1 \choose p}(-1)^{p}c_{p}\xi^p + O(1),
  \qquad N\to\infty,
\end{equation}
where $c_p={2p \choose p}/(p+1)$ are the Catalan numbers. Alternatively, they
can be encoded in the generating function \cite{jmp37pwb1996}
\begin{equation}
  \label{gene}
  G(s)=\sum_{m=1}^\infty M_m s^{m-1}=\frac{N}{2s}\left(\sqrt{1+\frac{4\xi s}
      {1-s}}-1\right).
\end{equation}
This is the function we shall obtain semiclassically.

\section{The semiclassical approximation}

A semiclassical approximation to the $t_{io}$ element of the transmission
matrix is available from the corresponding theory for the Green's function
\cite{c3hub1993}. It is given as a sum over all trajectories connecting
incoming channel $i$ to outgoing channel $o$, \begin{equation}\label{trans}
t_{io}=\sum_{\gamma:i\to o} A_\gamma e^{\i S_\gamma /\hbar},\end{equation}
where as usual $S_\gamma$ is the action of trajectory $\gamma$ and $A_\gamma$
is related to its stability. The semiclassical expression for the moments
contains $2m$ sums over classical trajectories, \begin{equation}\label{basic}
M^{\rm sc}_m=\sum_{\bm{i},\bm{o}} \sum_{\bg,\br} \prod_{j=1}^m
A_{\gamma_j}A^\ast_{\sigma_j}\left\langle
e^{i(S_{\bg}-S_{\br})/\hbar}\right\rangle.\end{equation} Here
$\bm{i}=\{i_1,\ldots,i_m\}$ and $\bm{o}=\{o_1,\ldots,o_m\}$ are sets of $m$
incoming and outgoing channels, respectively.

We are denoting by $\bg$ and $\br$ two sets of $m$ scattering trajectories:
$\gamma_j$ goes from $i_j$ to $o_j$ while $\sigma_j$ goes from $i_{j+1}$ to
$o_j$ ($i_{m+1}\equiv i_1$). The quantities $S_{\bg}=\sum_j S_ {\gamma_j}$ and
$S_{\br}=\sum_j S_{\sigma_j}$ are the total actions of these sets. The average
in (\ref{basic}) is taken around a classically small energy window, small
enough to keep the classical dynamics and the amplitudes $A_\gamma$ essentially
unchanged. The phase factor on the other hand is rapidly oscillating as
$\hbar\to 0$, and this selects from the sum only correlated trajectories, with
total action difference of order $\hbar$. We restrict ourselves to the leading
order in $N^{-1}$.

\begin{figure}[t]
\centerline{\includegraphics[clip,scale=1]{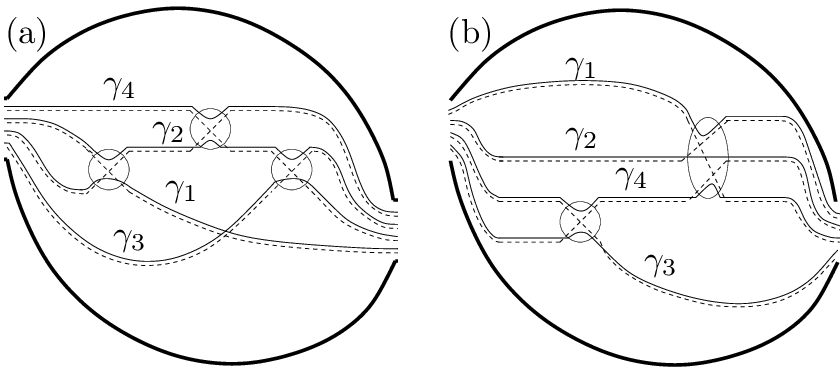}}
\caption{Schematic examples of correlated sets of classical
trajectories contributing to $M^{\rm sc}_4$. Each trajectory
$\gamma_j$ goes from incoming channel $i_j$ to outgoing channel
$o_j$, and is represented by a solid line. Trajectories $\sigma_j$,
which go from $i_{j+1}$ to $o_j$, are in dashed lines. The circles
mark the encounters, where trajectories switch partners (see text).}
\label{fig:corr}
\end{figure}

In what follows we identify the classical trajectories possessing the required
correlations and therefore contributing to (\ref{basic}). These are sets of
trajectories involving encounters, examples of which are shown in
Fig.\ref{fig:corr}. A pair $(\bg,\br)$ will contribute to (\ref{basic}) only if
there are places where $\gamma$-trajectories come very close to each other,
forming an ``encounter", and the corresponding $\sigma$-trajectories are
obtained by a process of `reconnection' at the encounters, such that $\sigma_j$
initially runs closely to $\gamma_{j+1}$ and ends up running closely to
$\gamma_j$. An encounter involving $\ell$ trajectories is called an
$\ell$-encounter. In between encounters the trajectories follow arcs, along
which the two sets are practically indistinguishable. The duration of a typical
encounter is of the order of the Ehrenfest time $\tau_E$, much smaller than the
typical duration of an arc which is proportional to the mean dwell time $\tau_D$. The
action difference between the two sets of trajectories comes almost entirely
from the vicinity of the encounters, and thus becomes small in the
semiclassical limit. This theory has been discussed in several previous
semiclassical calculations
\cite{prl89kr2002,prl95sh2005,prl91hs2003,prl96rsw2006,jpa39pb2006,kuipers,njp9sm2007}.

\begin{figure}[t]
  \label{fig:trees_ex}
  \centerline{\includegraphics[clip,scale=1.1]{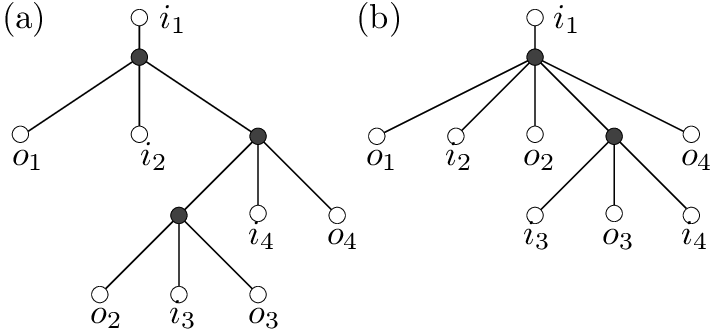}}
  \caption{The trees that correspond to Figure 1. The edges are the common
    arcs, the empty circles are the lead channels, and the shaded circles
    (``nodes'') are the encounters.  The edge emerging from $i_1$ is the root,
    and the channels are leaves.  Note that the leaves are ordered $i_1, o_1,
    i_2, \ldots, o_m$, starting with the root and going anti-clockwise. }
\end{figure}

To perform the calculation we must construct all possible sets $\bg$, and to this end we represent each
set by a diagram containing its `backbone' morphology of arcs and encounters.
The complicated arcs of the actual trajectories are represented by straight
edges; the encounters are represented by vertices of even degree
($\ell$-encounter has degree $2\ell$) and the lead channels are represented by
vertices of degree 1. The former vertices will be called {\em nodes\/} and
denoted by shaded circles, see Fig.~\ref{fig:trees_ex}.  The vertices of degree
1 will be called {\em leaves} and denoted by empty circles.  We will see that
our diagrams happen to be of a special kind, namely rooted planar trees.

To each of these diagrams we associate a vector $\v=(v_2,v_3,\ldots)$, where
$v_\ell$ is the number of $\ell$-nodes (or $\ell$-encounters). For example, one
of the graphs in Fig.2 has characteristic $\v=(3)$, and the other has
characteristic $\v=(1,1)$. If a diagram has characteristic $\v$, it contains
$V(\v)=\sum_\ell v_\ell$ nodes, while the number of edges is $L(\v)=m+\sum_\ell
\ell v_\ell$. Simple rules have been established to `read off' the contribution
of a given pair $(\bg,\br)$ to $M_m^{\rm sc}$. Their derivation involves
setting local coordinates at the encounters using Poincar\'e sections and invoking
ergodicity to write down a probability density for encounters leading to an
action difference $\Delta S$.  Then integrating over $\Delta S$, and the
possible duration of the arcs and applying a sum rule of Richter and Sieber
\cite{prl89kr2002}. We do not repeat this procedure, which has been reviewed in
detail in \cite{njp9sm2007}. The result is that {\it each arc contributes
$1/N$}, while {\it each encounter contributes $-N$}.  In the following sections
we will use these rules together with an enumeration of the contributing
diagrams to recover the random matrix prediction (\ref{rmt}) semiclassically.

\section{No coinciding channels}

Let us initially assume that all channels are distinct, and do some simple
power-counting with the channel number. There is a total of
$\sum_{\bm{i},\bm{o}}=N_1^{m}N_2^m$ possibilities for distributing the incoming
and outgoing channels. Suppose we have found a pair $(\bg,\br)$ represented by
a graph with characteristic $\v$. The edges produce a factor of $1/N^{L(\v)}$,
while the nodes produce $(-N)^{V(\v)}$. We thus have
$N_1^{m}N_2^{m}/N^{L(\v)-V(\v)}$. We want our result to be of leading order, so
we must maximize $\widetilde{V}(\v)-L(\v)$, where $\widetilde{V}=V+2m$ is the
total number of vertices (including encounters and channels). The quantity
$\widetilde{V}(\v)-L(\v)$ is the negative of the Euler characteristic of the
diagram, and it is well known that its maximal value is $1$. Our moments
therefore scale linearly with the number of channels, in agreement with
(\ref{rmt}). Moreover, it is also known that $\widetilde{V}-L=1$ if and only if
the diagram is a tree.

As the root of the tree we choose the edge containing $i_1$. The defining
feature of the diagram is the existence of a traversal $i_1\to o_1\to i_2 \to
\ldots \to o_m \to i_1$, such that each edge is traversed exactly twice: once
in each direction.  This implies that all leaves below any given node
(``below'' in the sense of the natural ordering with the root being on the top)
are consecutive with respect to the above traversal. Inductively, one can
conclude that the branches of the diagram can be arranged in such a way that
the leaves are ordered $i_1, o_1, i_2, \ldots, o_m$, starting with the root and
going anti-clockwise.  Conversely, any tree with $v_j$ nodes of degree $2j$ and
the leaves marked $i_1, o_1, i_2, \ldots, o_m$ anti-clockwise represents a
diagram.  A trajectory $\gamma_1$, for example, can be read off a tree by going
from $i_1$ to $o_1$ along the shortest path. Thus we have established that the
diagrams contributing to the leading order are in a one-to-one correspondence
with {\em planar\/} rooted trees.  The term ``planar'' refers to the fact that
the tree is defined up to an orientation-preserving homeomorphism of the plane
onto itself, i.e.\ swapping the branches (generally) changes the tree.

The total contribution of such a tree to $M_m^{\rm sc}$ is simply $(-1)^{V(\v)}
N\xi^m$. A tree with characteristic $\v$ has $L(\v)-V(\v)+1$ leaves so we
denote by $d(\v)=(L(\v)-V(\v)+1)/2$ the number of incoming/outgoing channels
associated with trees of characteristic $\v$. If $\mathcal{N}(\v)$ denotes the
number of trees characterized by $\v$, then their combined contribution is
$N\xi^mC_m$ with
\begin{equation}
  \label{eq:lead_term} C_m=\sum_{\v:\,d(\v)=m}\mathcal{N}(\v)(-1)^{V(\v)}.
\end{equation}
The function $\mathcal{N}(\v)$ was studied by Tutte \cite{tutte}, who found an
explicit formula for it,
\begin{equation}
  \label{eq:tutte}
  \mathcal{N}(\v) = \frac{(\widetilde{V}-2)!}{(2m-1)! \prod_{j} v_j!}.
\end{equation}
However, for the sake of introducing the machinery needed in the following
section, we take a different route and compute the above sum by making use of
generating functions. The main idea is to notice that if
\begin{equation}
  \label{eq:gx}
  f(x_2, x_3,\ldots) = \sum_{\v} \mathcal{N}(\v) x_2^{v_2} x_3^{v_3} \ldots
\end{equation}
is the generating function of $\mathcal{N}(\v)$ then $g(r)$, obtained by
setting $x_{n+1}=-r^n$, is the generating function of the numbers $C_m$,
\begin{equation}\label{eqg}
  g(r) = f(-r,-r^2,\ldots) = \sum_\v \mathcal{N}(\v)(-1)^{V(\v)}r^{d(\v)}
  = \sum_{m=1}^\infty C_{m}r^{m-1}.
\end{equation}

A tree characterized by $\v$ contains subtrees emerging from the top
$(n+1)$-node, which may be characterized by their own node vectors
$\v_1,\dots,\v_{2n+1}$, numbering left to right. One can therefore count the
number of trees characterized by $\v$ by counting all possible subtrees. This
implies a recursion relation for $\mathcal{N}(\v)$ (see \ref{ap:tree
recursion} for details),
\begin{equation}
  \label{eq:recur_N}
  \mathcal{N}(\v) = \sum_{n\ge 1} \sum_{\v_1\cdots\v_{2n+1}}
  \prod_{j=1}^{2n+1}\mathcal{N}(\v_j)\delta_{{\mathbf w},\v-\e_{n+1}},
\end{equation}
where ${\mathbf w}=\sum_j\v_j$ and $\e_n$ has $1$ in $n$-th entry and zero
everywhere else. Substituting the recursion relation in Eq.~(\ref{eq:gx}) we
see that $f$ satisfies $f=1+x_2 f^3+x_3f^5+\ldots$. Correspondingly
$g=1-rg^3-r^2g^5-\ldots$. Summing the geometric series we arrive at
$g=1-rg^2$, and thus \begin{equation} g(r)=\frac{-1+\sqrt{1+4r}}{2r}.\end{equation} When compared with
the generating function for the Catalan numbers $c_m$ this gives \begin{equation}\label{cm}
C_{m}=(-1)^{m-1}c_{m-1},\end{equation} in agreement with Eq.~(\ref{rmt}).

\section{Coinciding channels}

\begin{figure}[t]
\centerline{\includegraphics[clip]{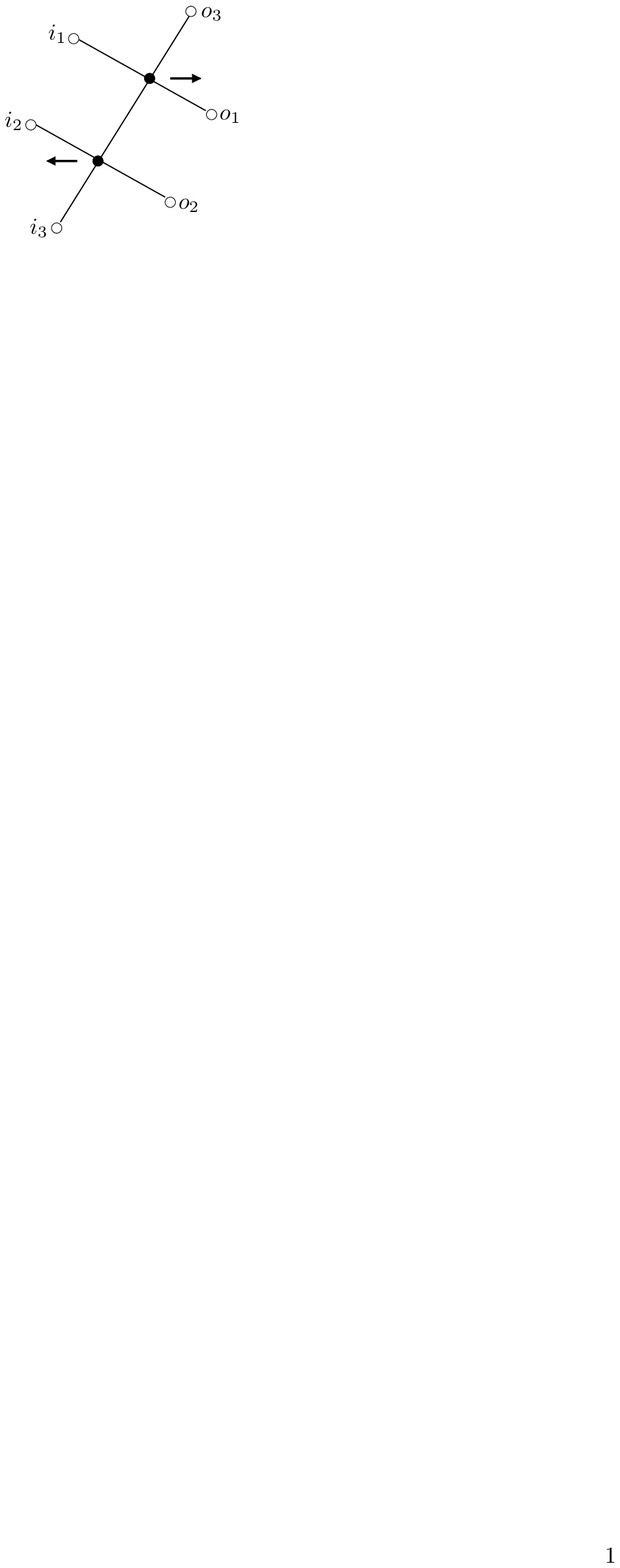}\includegraphics[clip]{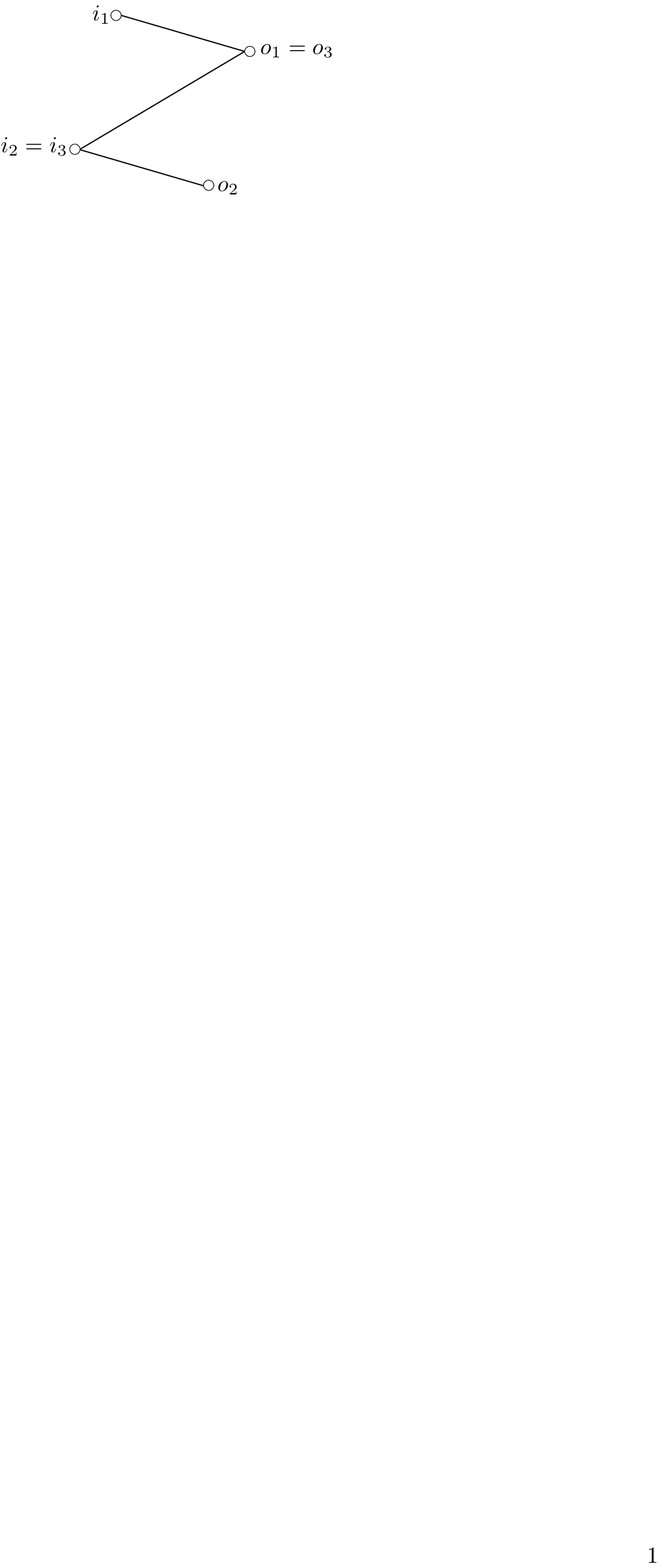}}
\caption{The diagram on the right, which has coinciding channels, is seen as a limiting case of the more
general diagram on the left. The encounters are continuously brought up to the leads, and eventually disappear
as all involved trajectories enter (or leave) from the same channel.}
  \label{fig:limit}
\end{figure}

The previous calculations assumed all channels to be different. If it happens
that $i_j=i_{j+1}$, then $\gamma_j$ and $\gamma_{j+1}$ enter the cavity from
the same channel. It is useful to view the corresponding diagram as arising
from a more general one (with $i_j\neq i_{j+1}$), in the limit when a
particular encounter happens closer and closer to one of the leads. After
taking this formal limit, all trajectories previously involved in the encounter
now enter from the same channel, and the encounter has disappeared. This is
illustrated in Fig.\ref{fig:limit} for a simple example with $m=3$. We are
actually neglecting any spatial dimension the encounter may have and treating
it as point-like. This is justified insofar as we consider $\tau_E\ll\tau_D$,
since the length and width of encounters are typically proportional to the
Ehrenfest time.

\begin{figure}[t]
\centerline{\includegraphics[scale=1,clip]{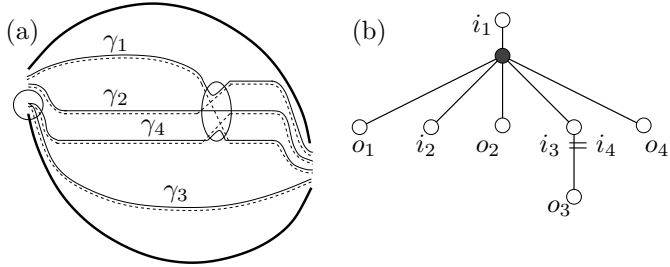}} \caption{Left:
Schematic of trajectories contributing to $M^{\rm sc}_4$ when
$i_3=i_4$. The crossing `touches' the incoming lead. Right: The
corresponding tree. Compared to the tree in Fig.2(b) the number of
segments is reduced by two, and the number of crossings is reduced
by one. }
  \label{fig:touch}
\end{figure}

In the situation just described we say that the node ``touches" the lead, which
may be incoming or outgoing. In any case, it no longer gives the usual $-N$
contribution. We now have to count in how many ways a given one of our trees
can have nodes touching leads. Consider for example the tree in Fig.2(a). The
lower node has two edges leading to outgoing leaves. This node could touch the
outgoing lead, and hence these edges would actually vanish. The top node could
touch the incoming lead, while the middle node cannot be made to touch any of
the leads. In Fig.\ref{fig:touch} we show the schematics and the corresponding
tree of a contribution similar to Fig.\ref{fig:corr}(b) but with $i_3=i_4$.

When an $(n+1)$-node touches an incoming lead, the number of edges is reduced
by $n+1$. The number of nodes is reduced by $1$, so the contribution gets
multiplied by $-N^n$. Because the number of channels involved in the sum is
also reduced, there is a factor of $N_1^{-n}$. If we define $z_1=N/N_1$, making
an $(n+1)$-node touch the incoming lead produces a factor $-z_1^n$ multiplying
the contribution of the tree. For a given tree $T$, denote by $q_{1,n}(T)$ the number
of $(n+1)$-nodes that can be made to touch the incoming lead. There are
${q_{1,n} \choose k}$ ways to have $k$ out of the $q_{1,n}$ such nodes actually
touch the lead. Therefore, the basic contribution must be multiplied by
\be\sum_{k=0}^{q_{1,n}} (-z_1^n)^k{q_{1,n} \choose k} =(1-z_1^n)^{q_{1,n}}.\ee
It is easy to see that each and every diagram with coinciding channels can be
derived from one and only one ``parent" diagram, which does not have any
coinciding channels. The conclusion is that we may consider only these
``parent" diagrams, provided we multiply each one of them by
$(1-z_1^n)^{q_{1,n}}(1-z_2^n)^{q_{2,n}}$, where the subscript $2$ refers to the
outgoing lead. This accounts for all possible contributions in which channels
coincide.

The above arguments amount to saying that, similarly to (\ref{eq:lead_term}),
the complete semiclassical moments are given by
\begin{equation}
  M^{\rm sc}_m (\xi)= N \xi^m \sum_{\v:\,d(\v)=m}(-1)^{V(\v)}P(z_1,z_2;\v),
\end{equation}
where the sum is over all characteristic vectors $\v$. Instead of including only the
number $\mathcal{N}(\v)$, as in (\ref{eq:lead_term}), we must instead use
\begin{equation}
  P(z_1,z_2;\v)  = \sum_{T\in \mathcal{T}_\v}\prod_{n\ge 1}
  (1-z_1^n)^{q_{1,n}(T)}(1-z_2^n)^{q_{2,n}(T)},
\end{equation}
where the sum is over $\mathcal{T}_\v$ the set of all trees characterized by $\v$. Notice that
$z_1z_2=z_1+z_2=\xi^{-1}$. Now we define the generating function of $P$,
\begin{equation} F(z_1, z_2;x_2,x_3, \ldots)
=\sum_{\v}P(z_1,z_2;\v)x_2^{v_2}x_3^{v_3}\ldots,\end{equation} and, by defining
$G^{\rm sc}(r)=N\xi F(z_1, z_2;-r,-r^2,\ldots)$, we obtain the analogue of
(\ref{eqg}) as \begin{equation}\label{bigG} G^{\rm sc}(r)=N\xi\sum_{m=1}^\infty
\frac{M^{\rm sc}_m(\xi)}{N\xi^m} r^{m-1}=\sum_{m=1}^\infty M^{\rm sc}_m(\xi)
\left(\frac{r}{\xi}\right)^{m-1}.\end{equation}

Unfortunately, it is not straightforward to write a recursion for $P$. This is
because, unlike any other node in the tree, the top node can be made to touch
{\em both} leads (although not at the same time). To circumvent this problem,
we define an auxiliary function,
\begin{equation}\P(z_1,z_2;\v) = \sum_{T\in \mathcal{T}_\v}\prod_{n\ge 1}
(1-z_1^n)^{q'_{1,n}(T)}(1-z_2^n)^{q_{2,n}(T)}, \end{equation} where $q'_{1,n}$
is defined as $q_{1,n}$ but excluding the top node. For $\P$ there is a natural
recursion relation, analogous to (\ref{eq:recur_N}), explained in the appendix.
It is given by
\begin{equation}\label{tau}\eqalign{\fl
  \P(z_1,z_2;\v) = \sum_{n\ge 1} \sum_{\v_1\cdots\v_{2n+1}}\P(z_1,z_2;\v_1)
  \cr\times\prod_{j=1}^{n}\P(z_2,z_1;\v_{2j})\P(z_1,z_2;\v_{2j+1})
  (1-z_2^n\, \delta_{\vu,\bm{0}}) \,\delta_{{\mathbf w},\v-\e_{n+1}},}
\end{equation}
where again we use $\mathbf{w}=\sum_{j=0}^{2n+1} \v_j$ and
$\vu=\sum_{j=0}^n\v_{2j+1}$. The sum is over the valency of the top node and
the characteristic vectors of the $2n+1$ subtrees that emanate from it. The
factor $z_2^n\, \delta_{\vu,\bm{0}}$ includes the contribution due to the top
node touching the outgoing lead which is only possible when all odd $\v_j$'s
vanish. The function $\P$ is useful because it is related to $P$ according to
\begin{equation}\label{TtoTau}\eqalign{\fl
  P(z_1,z_2;\v) = \sum_{n\ge 1} \sum_{\v_1\cdots\v_{2n+1}}
  \P(z_1,z_2;\v_1)\prod_{j=1}^{n}\P(z_2,z_1;\v_{2j})\cr\times\P(z_1,z_2;\v_{2j+1})
  (1-z_1^n\, \delta_{{\mathbf w},\vu}-z_2^n\, \delta_{\vu,\bm{0}})\, \delta_{{\mathbf w},\v-\e_{n+1}}\ .}
\end{equation}
Which now includes the contribution due to the top node touching the incoming lead
when all the even $\v_j$'s vanish. The details are also left to the appendix.

We now denote by $f(z_1, z_2;\x)$ the generating function of $\P$. The
recursion relation (\ref{tau}) implies \begin{equation}\label{eqf}
f(z_1,z_2;\x) = 1 + \sum_{n\ge 1}x_{n+1}(f^{n+1}(z_1,
z_2;\x)-z_2^n)f^n(z_2,z_1;\x).\end{equation} We again identify $x_{n+1}=-r^n$
and write $g(z_1,z_2;r)=f(z_1,z_2;-r,-r^2,\ldots)$. Making use of the geometric
series, we can reduce (\ref{eqf}) to
\begin{equation}\label{ftoh} g(z_1,z_2;r)=1+r(z_2-1)h(z_1, z_2;r),\end{equation} where we have
defined a function which is symmetric with respect to the variables $z_1,z_2$,
namely $h(z_1, z_2;r)=g(z_1,z_2;r)g(z_2,z_1;r)$. Substituting (\ref{ftoh}) back
into $h$ leads to the algebraic equation \be\label{heq} \xi
h=\xi(1-rh)^2+rh,\ee which can be solved to give \begin{equation}\label{h}
h=\frac{\xi+2\xi r-r-\sqrt {(\xi-r)(4\xi r+\xi-r)}}{2\xi r^{2}}.\end{equation}
Having this solution we can compute $g(z_1,z_2;r)$. Because of (\ref{TtoTau}),
the function $f(z_1, z_2;\x)$ is related to $F(z_1, z_2;\x)$ simply by $F = f -
\sum_{n\ge 1} x_{n+1}z_1^nf^{n+1}$. This implies that \begin{equation} G^{\rm
sc}(r)=\frac{N\xi g(z_1,z_2;r)}{1-rz_1g(z_1,z_2;r)}.\end{equation}

Let us multiply and divide the above expression by $(1-rz_2g(z_2,z_1;r))$. We
may then use that \be
(1-rz_1g(z_1,z_2;r))(1-rz_2g(z_2,z_1;r))=\frac{\xi-r}{\xi},\ee by virtue of
(\ref{ftoh}) and the definition of $z_{1,2}$. On the other hand, using
(\ref{ftoh}) again \be g(z_1,z_2;r)(1-rz_2g(z_2,z_1;r))=1-rh.\ee
Finally, taking the explicit formula (\ref{h}) into account and writing
$s=r/\xi$ leads to
\begin{equation} \sum_{m=1}^\infty M^{\rm
sc}_m(\xi)s^{m-1}=\frac{N}{2s}\left(\sqrt{1+\frac{4\xi s}
{1-s}}-1\right),\end{equation} which is identical to (\ref{gene}). Therefore,
all semiclassical moments are indeed equal to the corresponding random matrix
theory predictions. From the semiclassical point of view, the fact that the
leading order result is the same with or without time-reversal symmetry stems
from the fact that, being trees, the contributing diagrams contain no loops (in
contrast to Richter-Sieber pairs \cite{prl89kr2002}, for example).

Obtaining higher-order terms in $N^{-1}$, which are necessary to describe
experiments in which a relatively small number of channels are involved, would
require the incorporation of more general sets of trajectories, for which the
corresponding diagrams are no longer trees. Computing the full perturbative
series seems to be a very demanding task, especially for systems with
time-reversal symmetry. We should mention that some exact RMT results (valid
for arbitrary channel numbers) have appeared recently \cite{Savin,Vivos}. In
particular, for broken time reversal symmetry the average values of ${\rm
Tr}\mathcal{T}^m$, $({\rm Tr}\mathcal{T}^m)^2$ and $({\rm Tr}\mathcal{T})^m$,
where $\mathcal{T}=tt^\dag$, were obtained for general $m$ by one of the
present authors \cite{Novaes}.

\ack We have profited from interesting discussions with R.~Whitney.  GB and JMH
were supported by NFS grant DMS-0604859. MN was supported by EPSRC.

\appendix
\section{Tree recursion relations}\label{ap:tree recursion}

\begin{figure}[htb]
\begin{center}
\includegraphics[width=5cm]{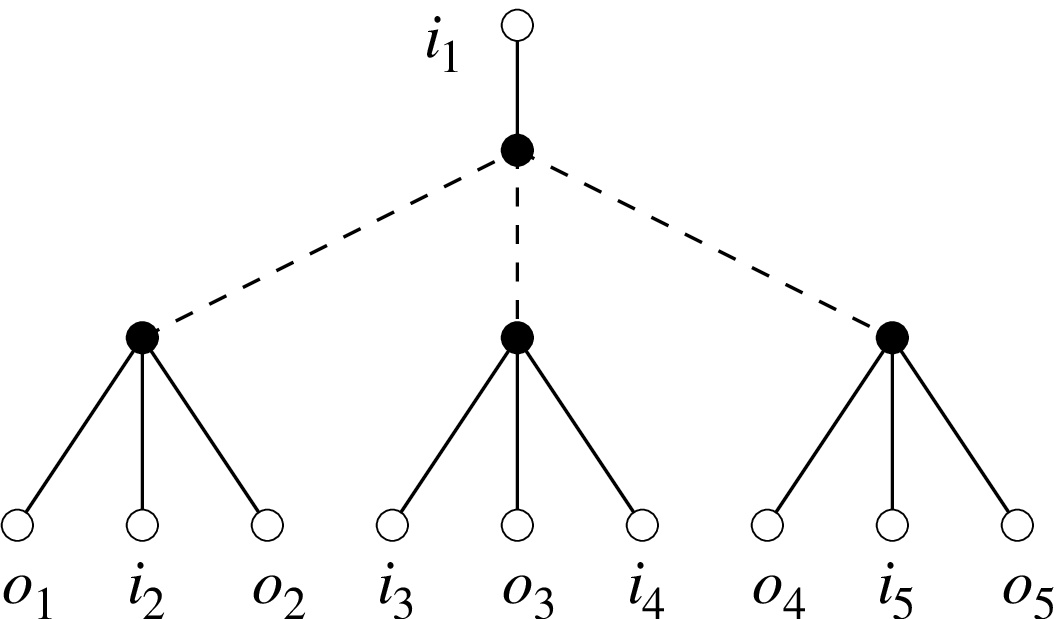}
\end{center}
\caption{A tree with characteristic $\v=(4)$ separates at the top node
 into three subtrees characterized by $\v_1=\v_2=\v_3=(1)$.}\label{fig:recur_ex}
\end{figure}

Let $T$ be a tree and $\mathcal{T}_\v$ the set of trees with characteristic $\v$.
We first establish the recurrence relation (\ref{eq:recur_N}) for $\mathcal{N}(\v)=|\mathcal{T}_\v|$,
the number of trees characterized by the node vector $\v$.
To derive the
relation we break the tree at the top $(n+1)$-node adjacent to the root.  The top
node has degree $2(n+1)$ and splitting the tree $T$ at this point the node becomes the root of $2n+1$
subtrees $T_1,\dots,T_{2n+1}$, characterized by vectors $\v_1,\dots,\v_{2n+1}$.  Clearly $\v=\sum_{j=1}^{2n+1}
\v_j +\e_n$ where $\e_n$ has $1$ in its $n$th entry and zero elsewhere,
representing the top node that was removed. Figure \ref{fig:recur_ex} shows a
tree with characteristic $(4)$.  This tree splits the at the top node, degree four ($n=1$),
into three subtrees each characterized by $(1)$. In general the number
of trees with top node degree $2(n+1)$ is given by the number of combinations
of subtrees, $\prod_{j=1}^{2n+1}\mathcal{N}(\v_j)$ where $\mathbf w=
\sum_{j=1}^{2n+1} \v_j = \v-\e_n$.  Summing over all allowed valencies of the
top node establishes the recursion relation,
\begin{equation}
  \label{eq:recur_N2}
  \mathcal{N}(\v) = \sum_{n\ge 1} \sum_{\v_1\cdots\v_{2n+1}}
  \prod_{j=1}^{2n+1}\mathcal{N}(\v_j)\delta_{{\mathbf w},\v-\e_{n+1}} \ .
\end{equation}

The recursion relation (\ref{tau}) for $\P$  can be generated in a similar
manner. To recap,
\begin{equation}\label{eq:Ptilde_defn}
\P(z_1,z_2;\v) = \sum_{T \in \mathcal{T}_\v} \prod_{m\ge 1}
(1-z_1^m)^{q'_{1,m}(T)}(1-z_2^m)^{q_{2,m}(T)},
\end{equation}
where $q_{2,m}(T)$ is the number of ways an $(m+1)$-node can touch the outgoing
lead and $q'_{1,m}(T)$ the number of $(m+1)$-nodes -- excluding the top node
adjacent to the root -- that can touch the incoming lead. For our trees the
root corresponds to the incoming lead and our definition of $\P$ excludes the
possibility of touching the root.

To establish the recursion relation we again consider breaking the tree into
subtrees $T_1, \dots , T_{2n+1}$ at the top $(n+1)$-node, numbering left to
right. The number of ways an $(m+1)$-node of the tree can touch the incoming
lead is
\begin{equation}\label{eq:q1_rec}
    q'_{1,m}(T)=q'_{1,m}(T_1)+\sum_{j=1}^{n} \big(q_{2,m} (T_{2j})+ q'_{1,m}(T_{2j+1}) \big) \ .
\end{equation}
This includes a change in the ordering of the leaves on the subtrees with even index.
On subtrees with odd index the first (left most) leaf must be an outgoing lead, labeled $o$, while the first lead of an even numbered subtree is labeled with the incoming lead $i$, see for example figure \ref{fig:recur_ex}.

\begin{figure}[htb]
\begin{center}
\includegraphics[width=8.5cm]{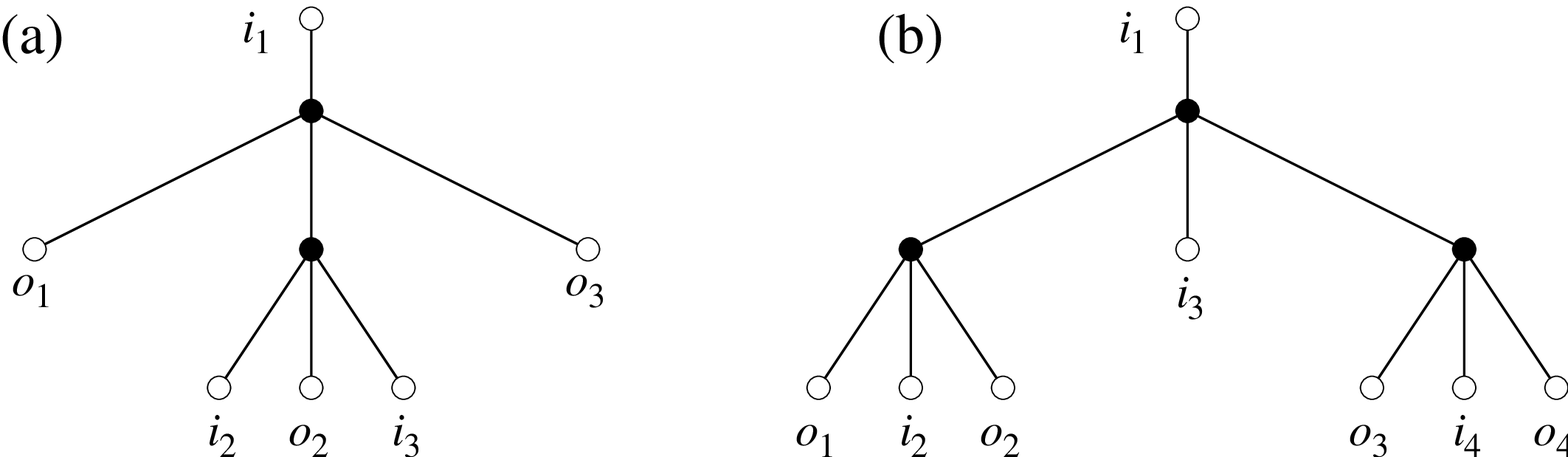}
\caption{(a) A tree with characteristic $\v=(2)$ where the top node can be made to touch the outgoing lead. (b) A tree with characteristic $\v=(3)$ where the top node can be made to touch the incoming lead.}\label{fig:top_node_contribution}
\end{center}
\end{figure}

To find the number of ways an $(m+1)$-node can touch the outgoing lead,
similarly one sums the $q_{2,m}$ of the odd subtrees and adds the $q_{1,m}'$ of
the even subtrees where the $o$ and $i$ leaf labels must be exchanged.  In
addition if $m=n$ the top node can contribute.  The top node may be made to
touch the outgoing lead if all the odd subtrees have characteristic $\bm{0}$,
i.e. every odd edge ends in a leaf. Figure \ref{fig:top_node_contribution}(a)
shows a tree where the top node can touch the outgoing lead. Therefore,
\begin{equation}\label{eq:q2_rec}
   q_{2,m}(T)=q_{2,m}(T_1)+\sum_{j=1}^{n} \big(q'_{1,m} (T_{2j})+ q_{2,m}(T_{2j+1}) \big)
 + \delta_{\vu, \bm{0}}\, \delta_{m,n}
\end{equation}
where $\vu=\sum_{j=0}^n\v_{2j+1}$.
Consequently $\P(z_1,z_2;\v)$ is expressed in terms of functions
$\P(\cdot,\cdot;\v_j)$ generated by the subtrees,
\begin{equation}\label{tau2}\eqalign{\fl
  \P(z_1,z_2;\v) = \sum_{n\ge 1} \sum_{\v_1\cdots\v_{2n+1}}\P(z_1,z_2;\v_1)
  \cr\times\prod_{j=1}^{n}\P(z_2,z_1;\v_{2j})\P(z_1,z_2;\v_{2j+1})
  (1-z_2^n\, \delta_{\vu,\bm{0}}) \, \delta_{{\mathbf w},\v-\e_{n+1}} }
\end{equation}

From $\P$ we recover $P$ by calculating the contribution generated when the top
node touches the incoming lead independently.  Recall the definition of $P$,
\begin{equation}\label{eq:P_defn}
  P(z_1,z_2;\v)  = \sum_{T\in \mathcal{T}_\v}\prod_{m\ge 1}
  (1-z_1^m)^{q_{1,m}(T)}(1-z_2^m)^{q_{2,m}(T)} \ .
\end{equation}
Comparing this with the definition of $\P$ (\ref{eq:Ptilde_defn}), we see that
terms in the sum are identical except when the top $(n+1)$-node of $T$ can
touch the incoming lead, in which case $q_{1,n}(T)=q_{1,n}'(T)+1$.  If we let
$\mathcal{R_\v}\subset \mathcal{T}_\v$ denote the set of trees that can touch
the root the definition of $P$ can be rewritten using $\P$.
\begin{equation}\label{eq:P=Ptilde-}
 \fl  P(z_1,z_2;\v)=\P(z_1,z_2;\v) - \sum_{T\in \mathcal{R_\v}}z_1^n
 \prod_{m\ge 1}
(1-z_1^m)^{q'_{1,m}(T)}(1-z_2^m)^{q_{2,m}(T)} \ ,
\end{equation}
where $n$ is determined by the degree of the node adjacent to the root. A tree
is in $\mathcal{R_\v}$ if, counting left to right, the even branches of the top
node all end in leaves, $\v_{2j}=\bm{0}$ for $j\in \{1,\dots, n\}$ or
equivalently ${\mathbf w}=\vu$.  Figure \ref{fig:top_node_contribution}(b)
shows a tree in $\mathcal{R_\v}$.  In equation (\ref{eq:P=Ptilde-}) the product
inside the sum over $\mathcal{R_\v}$ is the same as that appearing in the
definition of $\P$ and we can again break the trees in $\mathcal{R_\v}$ at the
top node to write the contribution in terms of subtrees.  Combining this with
the recursion relation (\ref{tau2}) for $\P$, we obtain the following formula
for $P$:
\begin{equation}\label{TtoTau2}\eqalign{\fl
  P(z_1,z_2;\v) =  \sum_{n\ge 1} \sum_{\v_1\cdots\v_{2n+1}}
  \P(z_1,z_2;\v_1) \prod_{j=1}^{n}\P(z_2,z_1;\v_{2j}) \cr \times \P(z_1,z_2;\v_{2j+1})
(1-z_1^n\, \delta_{{\mathbf w},\vu}-z_2^n\, \delta_{\vu,\bm{0}})\,
\delta_{{\mathbf w},\v-\e_{n+1}}}
\end{equation}

\end{document}